\documentclass[pra,amsmath,amssymb,reprint,showpacs]{revtex4-1}
\usepackage{amsmath, amssymb}
\usepackage{graphicx}
\usepackage{dcolumn}
\usepackage{bm}
\usepackage{color}

\newcommand{\sgn}{\operatorname{sgn}}

\begin{document}

\title{The Transition from Brownian Motion to Boom-and-Bust Dynamics \\
in Financial and Economic Systems}

\author{H. Lamba}
\affiliation{Department of Mathematical Sciences, George Mason University, Fairfax, USA}


\date{August 21st, 2012}

\begin{abstract}
Quasi-equilibrium models for aggregate variables are widely-used
throughout finance and
economics. The validity of such models depends 
crucially upon assuming that the systems'
participants behave both independently and in a Markovian fashion.


We present a simplified market model to demonstrate that 
herding effects between agents can cause a transition to
boom-and-bust dynamics at realistic parameter values. The model
can also be viewed as a novel stochastic particle system with
switching and reinjection. 



\end{abstract}

\maketitle

\section{Introduction}\label{sec:intro}
In the physical sciences using a stochastic differential equation
(SDE) to model the effect of exogenous noise upon an underlying
ODE is often straightforward. The noise consists of many
uncorrelated effects whose cumulative effect is well-approximated by a
Brownian process $B_s,\; s\geq 0$ and the ODE $df =
a(f,t)\;dt$ is replaced by an SDE $df = a(f,t)\; dt+ b(f,t) \; dB_t$.

However, in financial and socio-economic systems the inclusion of
exogenous noise (ie new information entering the system) is more
problematic --- even if the noise can 
be legitimately modeled as a Brownian process. This is because such
systems are themselves the aggregation of many individuals or trading
entities (referred to as {\em agents}) who typically \\
a) have differing
interpretations of the new information, \\ 
b) act differently depending  upon their own recent history (ie
non-Markovian behaviour), and \\
c) may not act
independently of each other.  

The standard approach in neoclassical economics and modern finance is
simply to `average away' these awkward effects by assuming the existence of a
single {\em representative agent} as in macroeconomics \cite{Kirman},
or by assuming 
  that the averaged reaction to new information is
  correct/rational, as in microeconomics and finance
  \cite{m61,f65}. In both cases, 
    the possibility of significant
  endogenous dynamics is removed from the models resulting in 
  unique, Markovian, (quasi)-equilibrium solutions.

\begin{figure}[!ht]
  \includegraphics[width=0.45\textwidth]{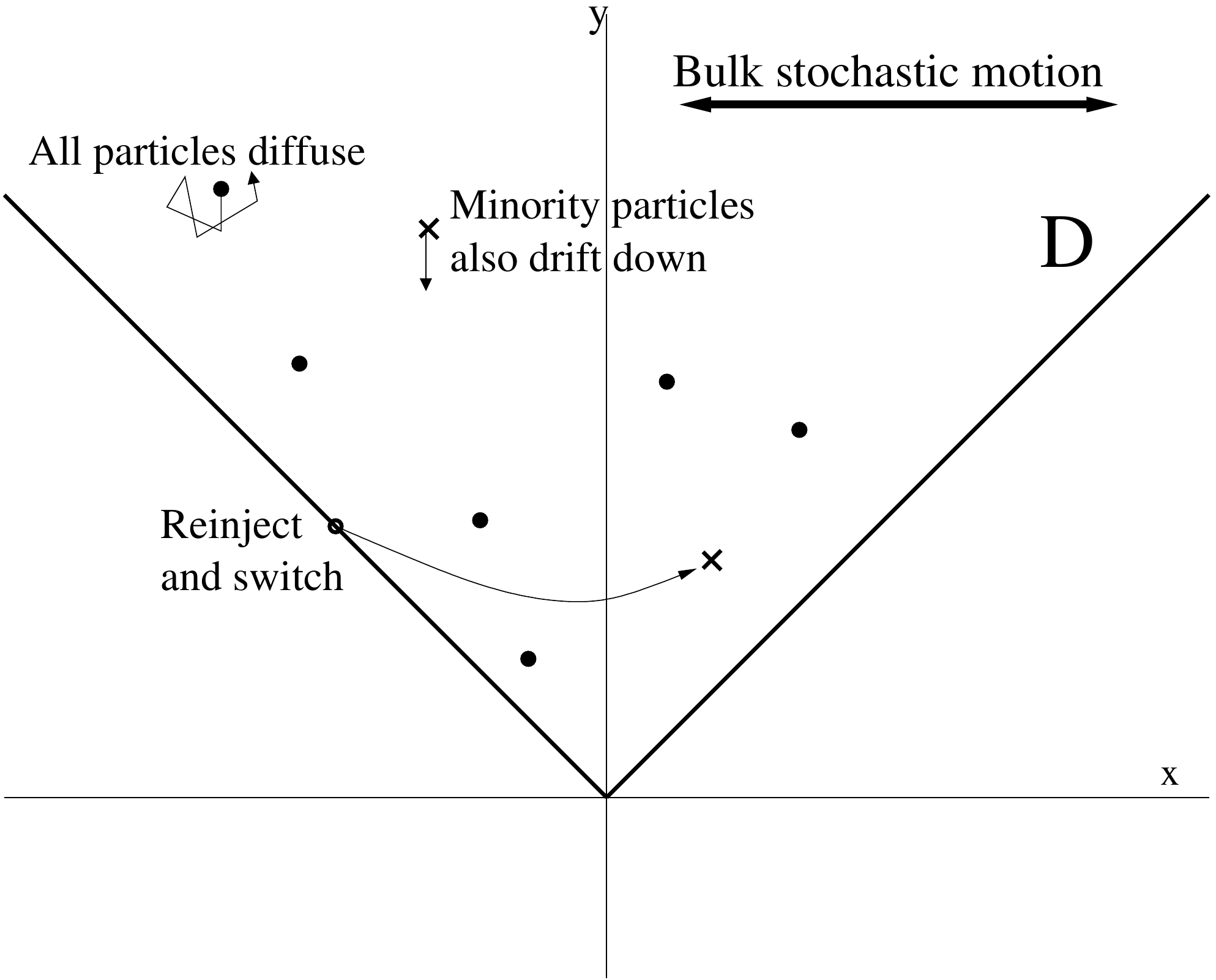} 
  \caption{All the $M$ signed particles are subject to a horizontal
    stochastic forcing and they also diffuse independently. Minority
    particles also drift downwards at a rate proportional to the
    imbalance. When a
    particle  hits the boundary it is reinjected with the opposite sign
    and a kick is added to the bulk forcing that can trigger a cascade
    (see text).}\label{fig:2d}
\end{figure}

In reaction to this, many Heterogeneous Agent Models (HAMs) have been
developed \cite{Hommes} that simulate the agents
directly. These have demonstrated that it is relatively easy to generate
aggregate output data, such as the price of a traded asset,
 that approximate reality better than the standard
averaging-type models. In particular the seemingly universal `stylized
facts' \cite{ms00,c01} of financial markets such as {\em heteroskedasticity}
(volatility clustering) and {\em leptokurtosis} (fat-tailed
price-return distributions resulting from booms-and-busts) have been
frequently reproduced. 
However, the effects of such research upon mainstream
modeling have been minimal perhaps, in part, because some HAMs require
fine tuning of important parameters, others are too complicated
to analyze, and the plethora of different HAMs means that many 
are mutually incompatible. 

The purpose of this report is to investigate the range of validity of the
quasi-equilibrium solutions 
obtained by ignoring endogenous effects such as a), b) and c) above.
We do this by introducing a simplified version of the modeling
framework introduced in \cite{ls08,l10} that  can also be
described as a particle system in two dimensions (Figure~\ref{fig:2d}).

\section{A stochastic particle system with reinjection and switching} \label{sec:model}

We define the open set $D \subset {\mathbb R}^2$ by $D=\{(x,y): -y
< x < y, \; y>0.$ There are $M$ signed particles (with states $+1$ or
$-1$) that move within $D$ subject to
three different motions. Firstly there is a bulk Brownian forcing
$B_t$ in
the $x$-direction that acts upon every particle. Secondly, each
particle has its own independent two-dimensional diffusion process.
Thirdly, for agents {\em in the minority state only}, there is a
downward (negative $y$-direction) drift that is proportional to the imbalance.  

When a particle hits the boundary $\partial D$ it is reinjected into
$D$ {\em with
the opposite sign}  according to some predefined probability
measure. Finally, when a particle does switch the position of the
other particles is kicked in the $x$-direction 
by a (small) amount $\pm \frac{2\kappa}{M}, \; \kappa
> 0,$ where the kick is positive if the switching particle goes from
the $-1$ state to $+1$ and negative if the switch is in the opposite
direction. 
Note that the particles do not interact locally or collide with one another\footnote{A web-based
interactive simulation of the model 
  can be found at http://math.gmu.edu/$\sim$harbir/PRLmarket.html}.

\subsection{Financial market interpretation}

We take as our starting point the standard geometric Brownian motion (gBm)
model of an asset price $p_t$ at time $t$ with $p_0=1$. It is more convenient to
use the log-price $r_t= \ln p_t$ which for constant drift $a$ and
volatility $b$ is given by the solution 
$r_t = at + b B_t$ to the SDE \begin{equation} dr_t = a\;dt + b\;
  dB_t. \label{gbm} \end{equation} 

Note that the solution $r_t$ depends only upon the value of
the exogenous Brownian process $B_t$ {\em at time $t$} and not upon
$\{B_s\}_{s=0}^t$. This seemingly trivial observation implies
that $r_t$ is Markovian and consistent with various notions of market
efficiency.
Thus gBm can be considered a paradigm for all
economic and financial models in which the aggregate variables are
in a quasi-equilibrium reacting adiabatically to new information.   

\begin{figure}[htp]
  \includegraphics[width=0.45\textwidth]{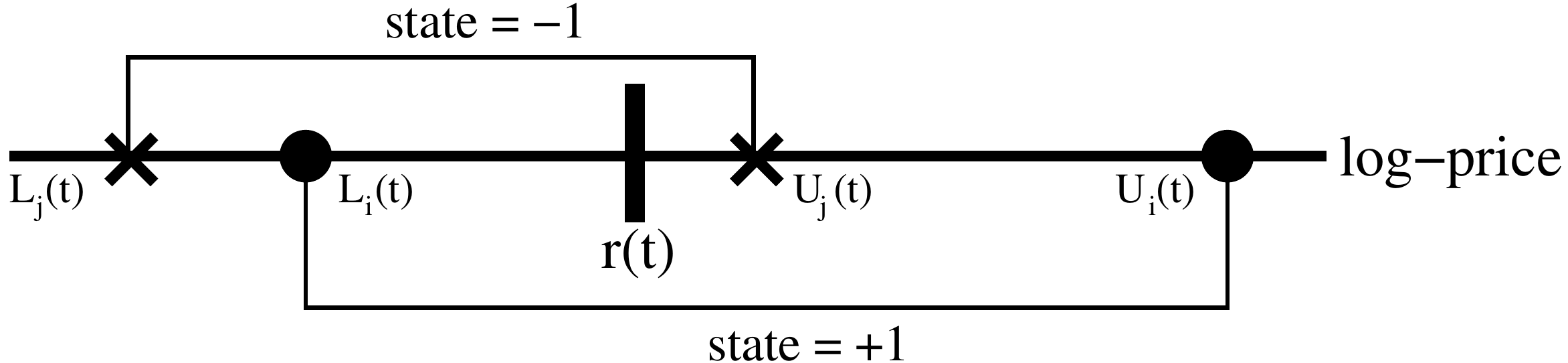} 
  \caption{A representation of the model showing two agents in
    opposite states. Agent $i$ is in the $+1$ state and is represented by the
  two circles at $L_i(t)$ and $U_i(t)$ while agent $j$ is in the $-1$
  state and is represented by the two crosses.}\label{fig:priceline}
\end{figure}

The instantaneous translation of new information $B_t$ into price changes is
effected by `fast' agents who will not be modeled directly. However,
we posit the existence of $M$ `slow' agents who are primarily
motivated by price changes rather than new information and act over
much longer timescales (weeks or months). 
At time $t$ the $i^{\rm
  th}$ slow agent is either in state $s_i(t)= +1$ (owning the asset)
or $s_i(t)= -1$
(not owning the asset) and the {\em sentiment} $\sigma(t) \in [-1,1]$ is
defined as $\sigma(t) = \sum_{i=1}^M s_i(t).$
The $i^{\rm th}$ slow agent is deemed to
have an evolving strategy that at time $t$ consists of an open interval
$(L_i(t),U_i(t))$ containing the  current log-price $r_t$ (see Figure~\ref{fig:priceline}). The $i^{\rm
  th}$ agent
switches state  whenever the price crosses either threshold, ie
$r_t=L_i(t)$ or $U_i(t)$,  and a new strategy interval is generated
straddling the current price.

We assume in addition that each threshold for every slow agent has its own
independent diffusion with rate $\alpha_i$ (corresponding to slow
agents' independently evolving
strategies) and those in the minority (whose state differs from
$\sgn(\sigma)$) also have their lower and upper thresholds drift
inwards each at a rate $C_i |\sigma|, \; C_i >0$. 

These {\em herding
  constants} $C_i$ are crucial as they provide the only (global)
coupling between agents. The inward drift of the minority agents' strategies
makes them more likely to switch to join the majority. Herding, and
other mimetic effects, appear to be a common feature of financial and
economic systems. Some causes are irrationally human while others may
be rational responses by, for example, fund managers not wishing to
deviate too far from the majority opinion and thereby risk
severely underperforming their average peer performance. The
reader is directed to \cite{l10} for a more 
detailed discussion of these and other modeling issues.

Finally, changes in the sentiment $\sigma$ feed back into the asset
price so that gBm (\ref{gbm}) is replaced with
\begin{equation} dr_t = a\;dt + b\;
  dB_t + \kappa \Delta \sigma \label{gbm+} \end{equation} 
where $\kappa > 0$ and the ratio $\kappa/b$ is a measure of the
relative impact upon $r_t$ of exogenous information versus
endogenous dynamics. Without loss of generality we let $a=0$ and $b=1$
by setting
the risk-free interest rate to zero and rescaling time.

One does not need to assume that all the slow agents are of equal
size, have equal strategy-diffusion, and equal herding
propensities. But if one does set $\alpha_i=\alpha$ and $C_i=C \;
\forall i$ then
one obtains the particle system simply by defining  the position of the
$i^{\rm th}$ particle as $(x_i,y_i) = (\frac{U_i+L_i}{2}-r_t,\frac{U_i-L_i}{2})$.
In other words, the bulk stochastic motion is due to exogenous noise;
the individual diffusions are caused by strategy-shifting of the slow
agents; the downward drift of minority agents is due to herding
effects; the reinjection and switching are the agents changing
investment position; and the kicks that occur at switches are
due to the change in sentiment affecting the asset price via a linear
supply/demand-price assumption.

\subsection{Limiting values of the parameters}

There are different parameter  limits that are potentially of interest.\\
1) \underline{$M \rightarrow \infty$} In the continuum limit the
particles are replaced by a pair of evolving density functions
$\rho^+(x,y,t)$ and $\rho^-(x,y,t)$ representing the density of each
agent state on $D$ --- such a mesoscopic Fokker-Planck description of
a related, but simpler, market
model can be found in \cite{glc12}. The presence of nonstandard boundary
conditions, global coupling, and bulk stochastic motion present
formidable analytic challenges for even the most basic questions of
existence and uniqueness of solutions. However, numerical simulations
strongly suggest that, minor discretization effects aside, the behaviour of
the system is independent of $M$  for $M \gtrapprox 1000$.\\
2) \underline{$B_t \rightarrow 0$} As the external information stream is
reduced the system settles into a state where $\sigma$  is close to
either $\pm  1$. Therefore this potentially useful simplification is
not available to us.\\
3) \underline{$\alpha \rightarrow 0$ or $\infty$} In the limit $\alpha
\rightarrow 0$ the particles do
not diffuse ie. the agents do not alter their thresholds between
trades/switches. This case was examined in \cite{cgls05} and the lack of
diffusion does not significantly change the boom-bust behaviour shown
below. On the other hand, for $\alpha \gg \max(1,C)$ the diffusion
dominates both the exogenous forcing and the herding/drifting and
equilibrium-type dynamics is re-established. This
case is unlikely in practice since slow agents will
alter their strategies more slowly than changes in the price of the asset.  \\
4) \underline{$C \rightarrow 0$} This limit is the focus of
the report. When $C=0$ the particles are uncoupled and if the system
is started with approximately equal distributions of $\pm 1 $ states
then $\sigma$ remains close to 0. Thus (\ref{gbm+}) reduces to
(\ref{gbm}) and the particle system becomes a standard equilibrium
model --- agents have differing expectations about the future which
causes them to trade but on average the price remains `efficient'
\cite{}. In Section~\ref{sec:num} we shall observe that endogenous
dynamics arise as $C$ is increased and the equilibrium solution is no
longer stable.\\
5) \underline{$\kappa \rightarrow 0$} For $\kappa >0$ one
agent switching can cause an avalanche of similar switches, especially
when the system is highly one-sided with $|\sigma|$ close to 1. 
 When $\kappa =0$ the particles
no longer provides kicks (or affect the price) when they switch
 although they are still coupled via
$C>0$. The sentiment $\sigma$ can still drift between $-1$ and $+1$
over long timescales but switching avalanches and large, sudden, price
changes do not occur. 

\section{Parameter estimation, Numerical Simulations and instability
}\label{sec:num}

\begin{figure}[htp]
  \includegraphics[width=0.5\textwidth]{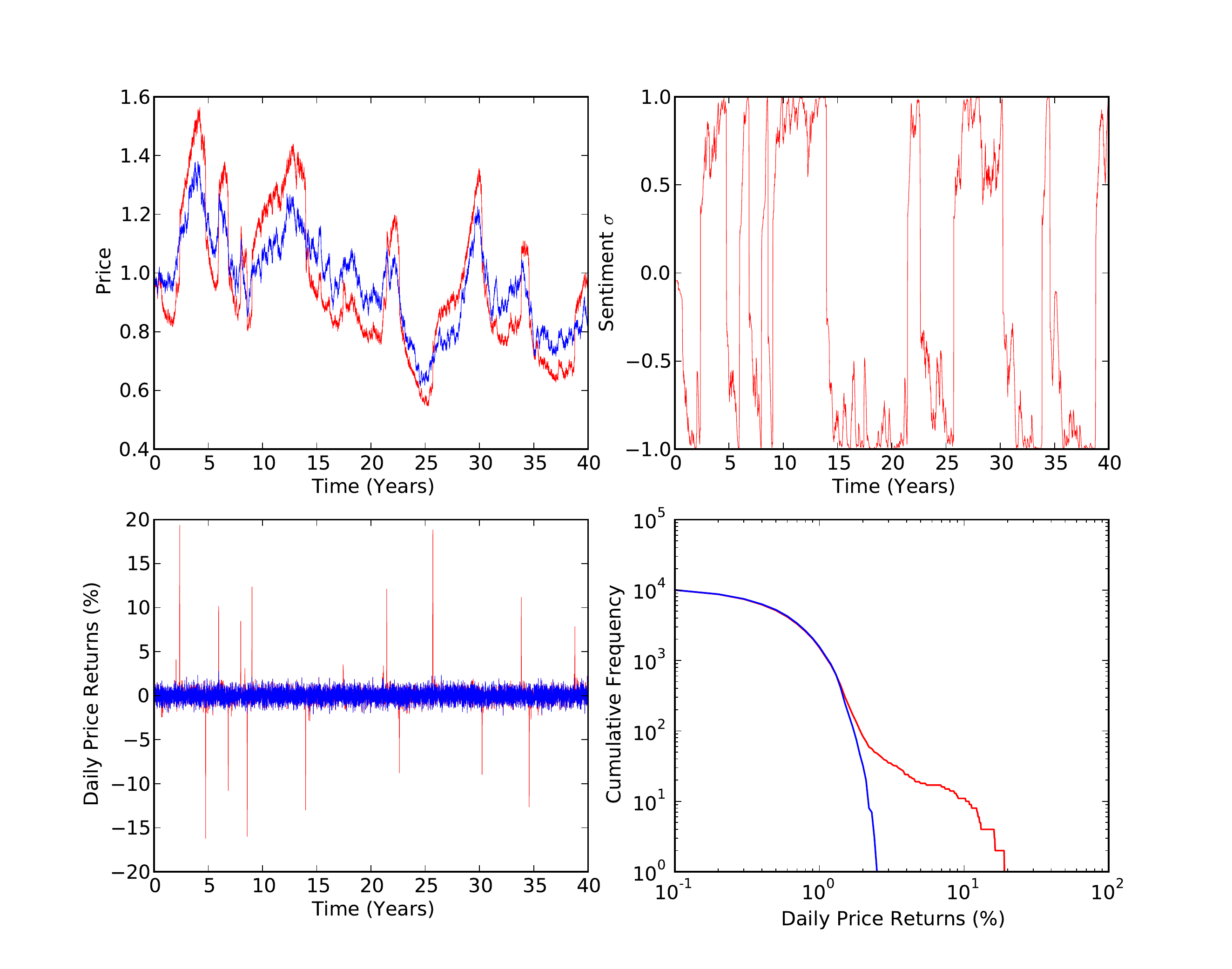} 
  \caption{In each picture, red corresponds to output with the
    parameters used in the text while blue represents outputs 
    for the gBm pricing model \protect{(\ref{gbm})} with the
    same $B_s$. 
    The top left figure shows the prices $p(t)$; top right is
    sentiment vs time; bottom left plots the daily price changes; and
    bottom right shows the cumulative log-log plot of daily price
    changes that exceed a given percentage.}\label{fig:testfig}
\end{figure}

In all the simulations below we use $M=1000$ and discretize 
using a timestep $h=0.000004$ which corresponds to approximately  1/10 of
a trading day if one assumes a daily standard deviation in prices of
$\approx 0.6\%$ due to new information. The price changes  of 10
consecutive timesteps are then summed to give
daily price return data making the difference between synchronous vs
asynchronous updating relatively unimportant.

We choose $\alpha = 0.2$ so that slow agents' strategies diffuse less strongly than
the price does. A conservative choice of $\kappa=0.2$ means that the difference in
price between neutral ($\sigma = 0$) and polarized markets $\sigma =
\pm 1$ is, from (\ref{gbm+}),  $\exp(0.2) \approx 22\%$.

After switching, an agent's thresholds are chosen randomly from a
Uniform distribution to be within 5\% and 25\% higher and lower than
the current price. This allows us to estimate $\kappa$ by supposing
that in a moderately polarized market with $|\sigma| = 0.5$ a typical
minority agent (outnumbered 3--1) would switch due to herding pressure
after approximately 
80 trading days (or 3 months, a typical reporting period for
investment performance)\cite{ss90}. 
The calculation $80 C |\sigma| = |\ln(0.85)|/0.00004$ gives $C \approx
100$. Finally, we note  that no fine-tuning of the parameters is
required for the observations below.

Figure~\ref{fig:testfig} shows the results of a typical simulation,
started close to equilibrium with agents' states equally mixed and run
for 40 years. The difference in price history  between the above
parameters and the equilibrium gBm solution is shown in the top left. The sudden
market reversals and over-reactions can be seen more clearly in the
top right plot where the market sentiment undergoes sudden shifts due to
switching cascades. These result in price returns (bottom left) that could quite
easily bankrupt anyone using excessive financial leverage and gBm as
an asset pricing model! 
Finally in the
bottom right the number of days on which the magnitude of the price
change exceeds a given percentage is plotted on log-log axes.
It should be emphasized that this is a simplified
version of the market model in \cite{l10} and an extra parameter
that improves the statistical agreement with real price data (by inducing volatility
clustering) has been ignored.

To conclude we examine the stability of the equilibrium gBm solution
using the herding parameter $C$ as a bifurcation parameter.
In order to quantify the level of disequilibrium in the
system we record the maximum value of $|\sigma|$ ignoring the first 10 
years of the simulation (to remove any possible transient effects caused
by the initial conditions) and average over 20 runs each for 
values of $0 \leq C \leq 40$. All the other parameters and the initial conditions are kept unchanged. 

\begin{figure}[htp]
\includegraphics[width=0.5\textwidth]{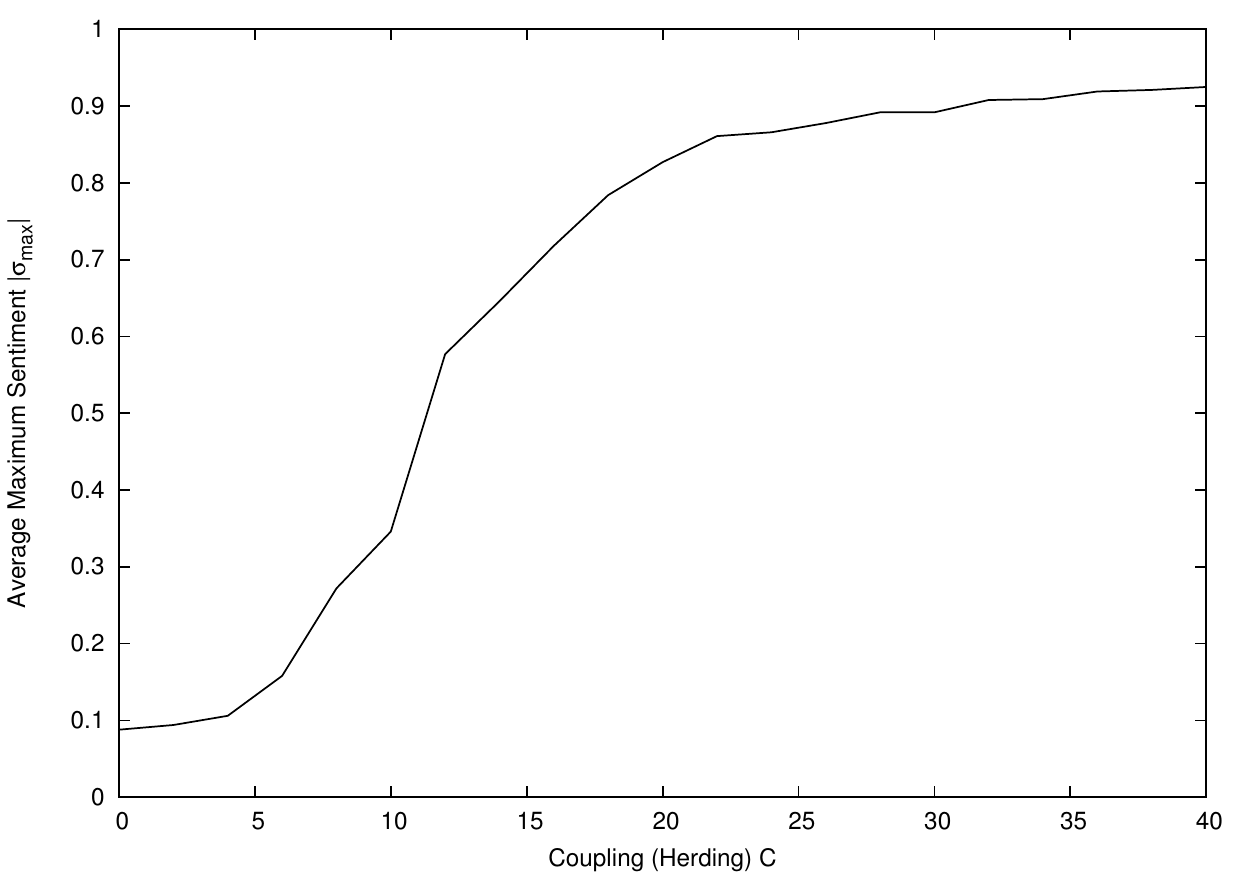}
  \caption{A measure of disequilibrium $|\sigma|_{\rm max}$
    averaged over 20 runs as the herding parameter $C$ changes.}
  \protect{\label{tab1}}
\end{figure}

The results in Figure~\ref{tab1} show that 
for values of $C$ as low as 20 the deviations from the equilibrium
solution are as large as the system will allow, with the large
majority of agents being in the same state at some point during the simulation. This is lower than 
the value of $C=100$ estimated above by a significant margin and it should be
noted that there are other phenomena, such as new investors and
money entering the asset market after a bubble has started, and
localized interactions between certain subsets of agents that might
cause further destabilization.

\section{Conclusions} \label{sec:conclusion}

Financial and economic systems are subject to many different kinds of
inter-dependence between agents and potential positive
feedbacks. However, even those mainstream models that attempt to
quantify such effects\cite{ss90} assume 
that the result will be  a shift of the equilibria to
nearby values without qualitatively changing the nature of the system.
However we have demonstrated that at least one
such form of coupling (herding) results in dis-equilibrium.
Furthermore the new dynamics occurs at realistic parameters and is
clearly recognizable as `boom-and-bust'. It is characterized by long periods of low-level
endogenous activity (long enough, certainly, to convince equilibrium-believers that the system is behaving
adiabatically) followed by large, sudden, reversals involving cascades of
switching agents triggered by price changes.

The model presented here is compatible with  existing
(non-mathematized) critiques of equilibrium
theory by Minsky and Soros\cite{m74,soros03}. Furthermore, work on
related models to
appear elsewhere shows that positive feedbacks can result in
similar non-equilibrium dynamics in more general 
micro and macro-economic situations. 

Finally, the model has
interesting links to other areas of mathematics and physics. In
\cite{l10} it was shown that the switching cascades can be described
using Queueing theory, with the price changes being equivalent to the
busy-period of a queue. The model also shares similarities with other
self-organizing systems such as the OFC earthquake
model\cite{ofc92}. And if one considers agents to be evolving hysteretic
operators then results concerning the interaction of stochastic
processes and hysteresis\cite{md05}  and phase transitions may provide valuable
insights  into the price dynamics.

The author thanks Michael Grinfeld and Rod Cross for numerous
enlightening conversations and Julian Todd for writing the browser-accessible
simulation of the particle system. The author also thanks Arjun
Sanghvi for his additional computations and the NSF for his support. 
%

\end{document}